\documentclass[onecolumn,showpacs,preprintnumbers,amsmath,amssymb]{revtex4}

\usepackage{graphicx}
\usepackage{dcolumn}
\usepackage{bm}

\begin{document}

\title{ The Dirac and  Gauge Yang-Mills Fields in Self-Consistent  Consideration}

\author{A.V.Koshelkin.}
 \altaffiliation[  koshelkin@mtu-net.ru; koshelkin@theor.mephi.ru; ]
 {}
\affiliation{Moscow Institute for Physics and Engineering,
Kashirskoye sh., 31, 115409 Moscow, Russia }
\date{\today}

\begin{abstract}
The  quasi-classical model in a gauge theory with the  Yang-Mills
(YM) field is developed. On a basis of the exact solution of the
Dirac equation in the $SU(N)$ gauge field, which is in the eikonal
approximation,  the Yang-Mills (YM) equations containing the
external fermion current are solved. The derived solutions are
quantized in the quasi-classical approach. The developed model
proves to have the self-consistent solutions of the Dirac and
Yang-Mills equations at $N\geq 3$. Thereat the solutions take
place provided that the fermion and  gauge fields exist
simultaneously, so that the fermion current completely compensates
the current generated by the gauge field due to it
self-interaction. The obtained solution are considered in the
context of QCD.

\end{abstract}

\pacs{11.15.-q, 03.70.+k, 11.10.-z }

\maketitle

\section{Introduction}

A study of  non-Abelian gauge fields plays an important  role in
the modern field theory\cite{1,2,3}. The non-Abelian gauge field
are a basis of  QCD\cite{4}. The knowledge of solutions of the YM
equations enable us to  understand features of  processes in the
strong interacting matter generated in collisions of heavy ions of
high energies\cite{5}. Primarily, this concerns studying the
observable states  of such  matter as well as the processes
accompanying evolution of the medium.

Studying the non-Abelian gauge fields has a very long history
which  started by the classic paper by C.N.Yang and
R.L.Mills\cite{6}. Since the paper\cite{6} has issued  a lot of
papers\cite{7,8,9,10,11,12,13,14,15,16,17} have been devoted to
deriving the solutions of  the YM equation in various situations.
The solution of the source-less YM equation in terms of plane
waves was derived in\cite{7,8}.  A wide class of solutions of the
YM  equation concerns (1+3) Minkowski space-time in the presence
of external sources\cite{9,10,11,12,13}. The YM equation   were
solved\cite{14} for the $SU(2)$ gauge field. The spherical
symmetric solutions are found for the $SU(2)$ fields in some
specific case of (1+2) space-time in Ref.\cite{15}. The Dirac
equation in the presence of the $SU(3)$ YM field is
considered\cite{16,17} in terms of the confinement problem. The
quark confinement in the curve space-time is studied in
Ref.\cite{18}. Rather detailed review of the paper devoting to
quantizing the YM field is in the monographs by A.Slavnov and
L.Faddev\cite{19}.

The consistent consideration  of the  strong interacting particles
(generated, for example, in collisions of high energy ions),
generally, demands solving the Dirac and Yang-Mills equations
simultaneously. The first step in studying such problem, naturally
(see Ref.\cite{20,21}) , is  an attempt to  derive the  solution
of these equations when the YM field has the form of some modified
plane wave so that  both the Dirac and Yang-Mills fields will be
in the confined region of space. The knowledge of the
self-consistent solution of the Yang-Mills and Dirac equations in
such approximation allows us to obtain the exact Green's function
of a fermion field. As  a result,  it enables to  drive both the
renormalized vertex functions and effective mass of a fermion as
well as to calculate the observable characteristics of the strong
interacting matter generated in collisions of high energy
ions\cite{5} beyond the perturbation theory.

In the present paper the  quasi-classical model in the $SU(N)$
gauge theory with the Yang Mills field is developed. The
self-consistent solutions of  both the  nonhomogeneous Yang-Mills
equation  and Dirac equations in an external field are derived
when the gauge Yang-Mills field is in the eikonal form. The
obtained solutions are quantized in the quasi-classical
approximation. It is shown that  the self-consistent solutions of
such equations take place when $N\geq 3$. They occur  provided
that the fermion and gauge fields simultaneously exist, so that a
fermion current completely compensates the current generated by
the gauge field due to it self-interaction. Thereat, there is no
energy flux from the range of space where the fields are
localized.

In the context of the multi particle problem in matter,  the
derived solutions mean that the Yang-Mills field is the modified
circularly-polarized wave which intensity depends strongly on
fermion density. In this way,  there are  both the individual and
collective states of fermi-particles in  matter. When the matter
is in equilibrium the type of the fermion states depends strongly
on such parameters as the temperature and density of the medium.
The fermion states appear to be one-particle at rather large
temperature. With decreasing the matter temperature they are
rearranged  so that the collective  states of the fermions
interacting with the YM field arise. Thereat, interaction between
the fermions and  YM field leads  to the re-normalization of a
fermion mass in the mean.  The re-normalized mass depends
significantly on the temperature of the matter.

The problem of the hadronization of an equilibrium quark-gluon
plasma is considered. In the case of the hadronization into the
lightest hadrons due to the phase transition of the first kind a
mass of the hadron is calculated. It appears to be of the order of
the  mass of a free pion.

The paper is organized as follows. Sections II consists of the
statement of the  problem and the  description of the  main
approaches. In Section III  the Yang-Mills equation is studied
when the gauge field has the form of the eikonal. Section IV is
devoted to deriving the exact solution of the Dirac equation in
the external $SU(N)$ gauge field being in the form of the eikonal
wave. The quantum consideration of the problem is developed in
Section V. The specifics of the Yang-Mills and fermion fields
obtained in the framework of the developed model  are studied in
Section VI. In the section VII the derived results are considered
in the context of QCD. The conclusion is the last section.
Appendix A contains the calculation of the operator exponent
appearing in solving the Dirac equation in an external non-Abelian
field. The detailed calculations of some important integrals are
in Appendix B.

\section{The YM equations in the presence of  external current }

We  consider the $SU(N)$ gauge field $A_a^{\nu}$ generated  by a
fermion current. It  satisfies  the following
equations\cite{19,22}:

\begin{widetext}
\begin{eqnarray}
&& \partial_{\mu} F^{\nu \mu}_a (x)  - g \cdot  f_{ab}^{\ \ c}
A_{\mu}^b (x) F_c^{\nu \mu} (x) = - g  {J_a}^{\nu} (x)
\\ \nonumber   \\ && F_a^{\nu \mu} (x) = \partial^{\nu} A_a^{\mu} (x) -
\partial^{\mu} A_a^{\nu} (x) - g\cdot f_{a}^{\ bc}A_b^{\nu} (x) A_c^{\mu} (x),
\\ \nonumber \\
&& {J_a}^{\nu} (x) = \  {\bar \Psi } (x) \gamma^{\nu}  T_a \Psi
(x) ,
\end{eqnarray}
\end{widetext}

where the  fermion  fields $\Psi (x), {\bar \Psi (x) }$ are
governed by the Dirac equation:

\begin{widetext}
\begin{eqnarray}
&& \left\{ i \gamma^{\mu} \left( \partial_{\mu} + i g \cdot
A_{\mu}^a (x) T_a \right) - m \right\} \Psi (x) = 0  \\ \nonumber  \\
&& {\bar \Psi} (x) \left\{ i \gamma^{\mu} \left(
{\overleftarrow\partial}_{\mu} - i g \cdot {A^{\ast}}_{\mu}^a (x)
T_a \right) + m \right\}  = 0; \ \ \ \ \ \ \ T_a = {1\over 2}
\lambda_a .
\end{eqnarray}
\end{widetext}
Here,  $m$ is a fermion mass, $g$ is the coupling constant;
$\gamma^{\nu}$ are the Dirac matrixes, $x\equiv x^\mu = (x^0 ;
{\vec x})$ is a vector in the Minkowski space-time;
$\partial_{\mu} = (\partial /\partial t ; \nabla )$; the Roman
letters numerate a basis in the space of the associated
representation of the $SU(N)$ group, so that $a,b,c = 1 \dots N^2
-1$. We use the signature $ diag \left( {\cal G}^{\mu \nu} \right)
= (1; -1; -1; -1)$ for the metric tensor ${\cal G }^{\mu \nu}$.
The line  and "dagger" over $\Psi$  mean the  Dirac  and hermitian
conjugation, respectively\cite{23}. Summing over any pair of the
repeated indexes is implied.

The symbols $T_a$ in Eqs.(3)-(5) are the generators of the $SU(N)$
group which satisfy the commutative relations and normalization
condition:

\begin{widetext}
\begin{eqnarray}
&& \left[ T_a, T_b \right]_-    =  T_a  T_b - T_b T_a =   i
f_{ab}^{\ \ c} T_c ; \ \ \ \ \ \ f_{ab}^{\ \ c}  = - 2\  i   \ Tr
\left( \left[ T_a, T_b \right]_- T_c \right)  \\
&& Tr \ ( T_a \ T_b )   ={1\over 2} \delta_{ab} ;
\end{eqnarray}
\end{widetext}
where $f_{ab}^{\ \ c}$ are the structure constant of the $SU(N)$
group, which are real and anti-symmetrical with respect to the
transposition in any pair of indexes;   $\delta_{ab}$ is the
Kroneker symbol. In the matrix representation the operators $( 2\
T_a )$ coincide with the Pauli and Gill-Mann matrixes when $N$ is
equal to $2$ or $3$, respectively.

It directly follows from Eqs.(6), (7) that

\begin{widetext}
\begin{eqnarray}
&& \left[ T_a, T_b \right]_+ = T_a  T_b + T_b T_a =  {1 \over  N}
\delta_{ab} + d_{abc} T^c , \ \ \ \ \ d_{abc} = 2 \ Tr \left(
\left[ T_a, T_b \right]_+ T_c \right)
\end{eqnarray}
\end{widetext}
where $d_{abc}$ is real and symmetrical with respect to the
transposition in any pair of indexes.

The main goal is to  derive the self-consistent solutions of  of
Eqs.(1)-(5) which will be localized in the confined region of
space. We  find the solution when  the field $A_{\nu}^a (x)$ is in
the form:

\begin{widetext}
\begin{eqnarray}
&&  A^{\nu}_a ( x )  = A^{\nu}_a (\varphi (x)) ,
\end{eqnarray}
\end{widetext}
where $\varphi (x)$ is some scalar function in the Minkowski
space-time which is  such that:

\begin{widetext}
\begin{eqnarray}
&&     (\partial_{\mu} \varphi ) (\partial^{\mu} \varphi ) \equiv
 k_\mu k^\mu = 0 ;
\end{eqnarray}
\end{widetext}
The last formula determines the well known eikonal approximation
where  $\varphi(x)$ can be interpretable as the function governing
the wave surface of the field $A^\nu_a$.

We take the  axial gauge   for the field $A_{\mu}^a (x)$ :

\begin{widetext}
\begin{eqnarray}
&& \partial^{\mu} A_{\mu}^a  = 0 ;  \ \ \ \ \ \ \ \ k^{\mu} {\dot
A}_{\mu}^a  = 0,
\end{eqnarray}
\end{widetext}
where the dot over the letter means  differentiation with respect
to  the introduced variable $\varphi $.

Taking into account of both the   dependence of  $A_{\nu}^a (x)$
on the variable $x$ via the function  $\varphi (x)$ and formulae
(10), (11), we derive from Eqs.(1), (2):

\begin{widetext}
\begin{eqnarray}&& 2 g k^{\mu} \  f_{ab}^{\ \ c}
A_{\mu}^b (\varphi) \dot{A}_c^{\nu} (\varphi)
 - \left( \partial_\mu \partial^\mu  \varphi (x) \right) \cdot {\dot {A}_a^{\nu} }  - g k^{\nu} \  f_{ab}^{\ \ c}
A_{\mu}^b (\varphi) \dot{A}_c^{\mu} (\varphi) +  g^2  f_{ab}^{\ \
c}\ f_{c}^{\ \ sr} \left\{ A_{\mu}^b (\varphi) A^{\nu}_s (\varphi)
A^{\mu}_r (\varphi)
 \right\} = - g\  J_a^\nu (x) ; \nonumber \\ \nonumber \\
 && J_a^\nu (x) = {\bar \Psi} ( x ) \gamma^\nu T_a \Psi ( x ) .   \
\end{eqnarray}
\end{widetext}

It follows from Eq.(12) that in order to derive the solution of
the YM equation it is necessary to calculate the  fermion current
$J_a^\nu (x)$, which, in its turn,  is governed by the solutions
of the Dirac equation in the external field $A_a^\nu (x) $.

To do it we assume that the field $A^{\nu}_a (\varphi)$ can
expanded as follows  in the local frame:

\begin{widetext}
\begin{eqnarray}
&&  A^{\nu}_a (\varphi)  =   A  \left(   e_{(1)}^\nu (\varphi)
\cos \left( \varphi (x)  + \varphi_a  \right)  +   \ e_{(2)}^\nu
(\varphi)  \sin \left( \varphi ( x ) + \varphi_a \right) \right) +
{\cal B}_a  \
\partial^\nu
\varphi (x) \, \nonumber \\
&&  e_{(1)}^\nu  {e_{(2)}}_\nu =  e_{(1)}^\nu k_\nu = e_{(2)}^\nu
k_\nu = 0 ; \ \ \ {\dot e_{(1)}^\nu } = e_{(2)}^\nu ; \ \ \  {\dot
e_{(2)}^\nu } = - e_{(1)}^\nu   ; \ \ \ \ \
 k^\nu
\equiv
\partial^\nu \varphi (x) ,
\end{eqnarray}
\end{widetext}
where    $e_{(1), (2)}^\nu (\varphi)$ are the space-like 4-vectors
on the wave surface $\varphi (x)$ which are independent on the
group variable $a$; the symbols $A$, ${\cal B}_a$ and $\varphi_a$
are some constants in the Minkowski space-time. They are
determined via the initial condition of the studied problem. It is
obvious that the function $\varphi (x)$ can be taken so that the
field $A_{\nu}^a (x)$ will be localized in the confined region of
space.

\section{Fermions in the external YM field}

To obtain the fermion field $\Psi (x)$ we go from Eq.(4) to the
so-called quadric Dirac equation which  has the following form:

\begin{widetext}
\begin{eqnarray}
&& \left\{ - \partial_\mu \partial^\mu - m^2 +g^2 \left(\gamma^\mu
A^a_\mu T_a \right)^2 + 2 i g \left(\gamma^\nu A^a_\nu T_a \right)
\left(\gamma^\mu \partial_\mu \right) + i g \left(\gamma^\nu k_\nu
 \right) \left(\gamma^\mu
{\dot A}^a_\mu T_a \right) \right\} \Phi (x) = 0 ; \nonumber \\
&& \Psi (x) = \left\{ { i \gamma^{\mu} \left( \partial_{\mu} - i g
\cdot A_{\mu}^a (x) T_a \right) + m \over 2 m} \right\} \Phi (x)
\end{eqnarray}
\end{widetext}

First, to derive the solution  of the last equation we simplify
the third term in the left-hand side of  Eq.(14).

Let  the initial  conditions be so that the phases $\varphi_a$ in
Eq.(13) satisfy the equations:

\begin{widetext}
\begin{eqnarray}
&& d_a^{\ bc} \ \cos (\varphi_a - \varphi_b ) = 0.
\end{eqnarray}
\end{widetext}

Then, using Eqs.(6)-(8) and  relations for the
$\gamma$-matrixes\cite{23,24} we obtain after  direct
calculations:

\begin{widetext}
\begin{eqnarray}
&& \left(\gamma^\mu A^a_\mu T_a \right)^2 = \left( {1\over 2 N}
\delta_{ab} + {1\over 2} d_{abc} T^c + {i\over 2} f_{abc}
T^c\right)  \left( {\cal G}^{\mu \nu } + \sigma^{\mu \nu} \right)
A_\mu^a A_\nu^b  = {1\over 2 N} A_\mu^a A^\mu_a \ ; \nonumber \\
&& \sigma^{\mu \nu} = {1\over 2} \left( \gamma^\mu \gamma^\nu -
\gamma^\nu \gamma^\mu \right)
\end{eqnarray}
\end{widetext}
We note,  that the term containing  $( f_{abc} T^c \sigma^{\mu
\nu} A_\mu^a A_\nu^b)  $ disappears since the vectors $e_{(1),
(2)}^\nu$, which are in the expansion  (13), are independent on
the group variable $a$. The same takes place with respect to all
terms containing  $\sigma^{\mu \nu}$  in Eq.(15).

Let us find the solution of Eq.(14) in the following  form:

\begin{widetext}
\begin{eqnarray}
&&  \Phi (x) \equiv \Phi_{\sigma, \alpha} (x, p) = e^{-ip x} \
\cdot F_{\sigma , \alpha} (\varphi ).
\end{eqnarray}
\end{widetext}
where $F_{\sigma, \alpha} (\varphi )$ is some multicomponent
function which is the generalized Dirac spinor. It depends on both
the spin variable $\sigma$ and the variable $\alpha$ which
specifies the state of a fermion in  the space of the fundamental
representation of the $ SU(N) $ group, thereat  $\alpha = 1 \div
N$;  \ $ p^\nu = \left( p^0 , {\vec p} \right)$ is some 4-vector.

We substitute  $\Phi_{a, \alpha} (x, p)$ given by  Eq.(17) into
the formula (14). Using  relations for the $\gamma$-
matrixes\cite{23,24},  independence of $e_{(1), (2)}^\nu$ on the
group variable $a$ in the local frame (see Eq.(13)) as well as
Eq.(10), (11), (16), we obtain:

\begin{widetext}
\begin{eqnarray}
&& \left\{ p^2 - m^2 - {g^2 (N^2 - 1 ) A^2  \over 2 N}  - 2 g
\left( T_a A_\mu^a p^\mu \right)\  - i g \left(\gamma^\nu k_\nu
 \right) \left( \gamma^\mu T_a
{\dot A}_\mu^a  \right) \right\} F_{\sigma , \alpha } (\varphi )+
 i\  ( p k ) \   {\dot F}_{\sigma
, \alpha } (\varphi ) = 0 ;
\nonumber \\
&& k_\mu = \partial_\mu  \varphi (x) ; \ \ \ \  \ ( p k) = p^\mu
k_\mu ;
\end{eqnarray}
\end{widetext}
where the dot over ${\dot F}_{\sigma , \alpha } (\varphi )$ means
 derivative with respect to the variable $\varphi $.

In obtaining the last equation we neglect $\vert \partial_\mu
k^\mu \vert$ as compared with $ \vert ( p k ) \vert$ (see
Eq.(18)). This means that the wave length $\lambda_{YM}$ of the YM
field  is unchangeable on the  scale which is of the order of  the
de Broglie wave length of a fermion $\lambda_F$:

\begin{widetext}
\begin{eqnarray}
&& \vert \partial_\nu k^\mu \vert \lesssim \vert \partial_\mu
k^\mu \vert = \vert
\partial_\mu
\partial^\mu \varphi (x) \vert \sim \left\vert  {d \lambda_{YM} \over \lambda^2_{YM}
\ d x} \right\vert \ll {1\over \lambda_{YM}\ \lambda_F} \sim \vert
(p k)\vert  \ \ \Leftrightarrow \ \ \left\vert  {d \lambda_{YM}
\over \ d x} \right\vert \ll {\lambda_{YM} \over \ \lambda_F} \ll
1
\end{eqnarray}
\end{widetext}
The last inequality corresponds  to the so called quasi-classical
approximation and means that $\partial_\mu  k^\mu  = 0$. The
condition $\partial_\mu  k^\mu  = 0$ can be treated as the scale
invariance of the wave surface of the YM field. Thereat, the form
of the wave surface is determined by  the  harmonic functions
satisfying D'Lambert equation, $\partial_\mu \partial^\mu \varphi
(x) = 0$ .

In the quasi-classical approximation governed by  the inequality
(19) we neglect the unexplicit dependence of $k^\mu$ on the
eikonal $\varphi (x)$ and assume that $k^\mu$  only varies along
the wave surface. Then, the solution of Eq.(18) can be written in
the form:

\begin{widetext}
\begin{eqnarray}
&& F_{\sigma , \alpha } (\varphi ) =  \exp \left(  - i g^2 {( N^2
- 1) A^2 \over  2 N ( p k ) } \right)\ \exp \left\{   - i g \ T_a
{ {i\over 2} \left(\gamma^\nu k_\nu
 \right) \left( \gamma^\mu
 A_\mu^a (\varphi ) \right) + \int\limits_0^\varphi d \varphi^\prime \left( A_\mu^a (\varphi^\prime ) p^\mu
\right)\   \over  ( p k ) } \right\} \ u_\sigma (p) \cdot
 v_\alpha , \nonumber \\
 &&  p^2 = m^2 ; \ \ \ \partial_\mu  k^\mu = \partial_\mu \partial^\mu \varphi (x) =
0 ;
\end{eqnarray}
\end{widetext}
where $u_\sigma (p)$ and $ v_\alpha$  are  spinors which are
elements in the space of the corresponding representations.

The second  exponent in Eq.(20) is the operator acting   on the
spinors $u_\sigma (p)$ and $ v_\alpha$. The transforming the
exponent in the same way as it has been done  in Ref.\cite{25}
(see also Appendix A), we obtain:

\begin{widetext}
\begin{eqnarray}
&& \exp \left\{   - i g \ T_a { {i\over 2} \left(\gamma^\nu k_\nu
 \right) \left( \gamma^\mu
 A_\mu^a (\varphi) \right) + \int\limits_0^\varphi d \varphi^\prime \left( A_\mu^a (\varphi^\prime ) p^\mu
\right)\   \over  ( p k ) } \right\}  =
 \nonumber \\
&&  \cos \theta \Bigg \{ \left( 1 -  i g  T_a  { \tan \theta \over
\theta ( p k ) }  \  \int\limits_0^\varphi d \varphi^\prime \left(
A_\mu^a p^\mu \right) \right) + {g \left( \gamma^\nu k_\nu \right)
\left( \gamma^\mu A_\mu^a \right) \over 2 ( p k ) } \cdot \Bigg[
{\tan \theta \over \theta}\ T_a  + \nonumber \\
&& {g\over  ( p k) } \ {1 \over 2 N} \  \left( - i {\tan \theta
\over \theta} + {g\over  ( p k)}  {\theta - \tan \theta \over
\theta^3} T_b \ \int\limits_0^\varphi d \varphi^\prime \left(
A_\mu^b p^\mu \right) \right) \int\limits_0^\varphi d
\varphi^\prime \left( A_\nu^a p^\nu \right) \Bigg] \Bigg\}; \nonumber \\
&& \theta =  { g \over  ( p k ) } \sqrt{{1 \over 2 N}} \left(
\int\limits_0^\varphi d \varphi^\prime \left( A_\mu^a (
\varphi^\prime ) \  p^\mu \right) \ \int\limits_0^\varphi d
\varphi^{\prime \prime} \left( A_a^\nu ( \varphi^{\prime \prime })
\ p_\nu \right) \right)^{1\over 2}
\end{eqnarray}
\end{widetext}

We  substitute the exponent given by the last formula into
Eq.(20). After that,
 using Eqs.(17) and (14) we obtain

\begin{widetext}
\begin{eqnarray}
&& \Psi_{\sigma, \alpha} (x, p) =  \Phi_{\sigma, \alpha} (x, p) =
\ \cos \theta \cdot \exp \left(  - i g^2 {( N^2 - 1) A^2 \over  2
N ( p k ) } \varphi - i p x \right)\  \Bigg \{ \left( 1 - i g T_a
{ \tan \theta \over \theta ( p k ) }  \ \int\limits_0^\varphi d
\varphi^\prime \left( A_\mu^a p^\mu \right) \right) + \nonumber
\\
&& {g \left( \gamma^\nu k_\nu \right) \left( \gamma^\mu A_\mu^a
\right) \over 2 ( p k ) } \cdot \Bigg[ {\tan \theta \over \theta}\
T_a  +  {g\over  ( p k) } \ {1\over 2N} \  \left( - i {\tan \theta
\over \theta} + {g\over  ( p k)}  {\theta - \tan \theta \over
\theta^3} T_b \ \int\limits_0^\varphi d \varphi^\prime \left(
A_\mu^b p^\mu \right) \right) \int\limits_0^\varphi d
\varphi^\prime \left( A_\nu^a p^\nu \right) \Bigg] \Bigg\}\
u_\sigma (p) \cdot
 v_\alpha ;  \nonumber \\ \nonumber \\
 && \theta =  { g \over  ( p k ) } \sqrt{{1\over 2N}} \left(
\int\limits_0^\varphi d \varphi^\prime \left( A_\mu^a (
\varphi^\prime ) \  p^\mu \right) \ \int\limits_0^\varphi d
\varphi^{\prime \prime}\left( A_a^\mu ( \varphi^{\prime \prime })
\ p_\mu \right) \right)^{1\over 2} ; \ \ \ \  (\partial_\nu\ k^\nu
) = (\partial_\nu\
\partial^\nu ) \varphi (x) = 0 .
\end{eqnarray}
\end{widetext}

\subsection {General Solution of Dirac Equation in External Field}

In order to derive the general solution of the Dirac equation we
need to specify the   physical sense of the spinors $u_\sigma (p),
v_\alpha $  and the 4-vector $p$ which are in Eq.(22).

First, we require  that  the wave function (22) coincides with the
solution of the free Dirac equation at $\varphi = 0$. This means
that the spinors $u_\sigma (p)$ satisfy the relations:

\begin{widetext}
\begin{eqnarray}
&&  \sigma^{\mu \nu } k_\mu A_\nu (\varphi = 0 )  \ u_\sigma (p) =
0 \ ;  \ \ \ \ \  {\bar u}_\sigma (p) u_\lambda (p^\prime ) = \pm
2m \ \delta_{\sigma \lambda} \ \delta_{p p^\prime} ; \ \ \ p^2 =m
^2 ,
\end{eqnarray}
\end{widetext}
where $u_\sigma (p)$ are the bispinors of the free Dirac field.
Thereat, the first relation in Eq.(23) fixes the fundamental set
of solutions of Eq.(14) which is determined by the parameter $p$.
The plus and minus signs in Eq.(23) correspond to the Dirac scalar
production of the spinors $u_\sigma (p)$ and $u_\sigma (- p)$,
respectively.

Let us  clarify the physical sense  of the 4-vector $p^\mu$
arising in Eq.(17). To do it  we consider the  projections of the
momentum operator, ${\hat p}^\mu = -i \partial^\mu$,   along  the
$k^\mu$-direction and on the plane which is perpendicular to the
vector $k^\mu$ in the local frame govern by Eq.(13). For
definiteness sake, we take  $k^\mu = (1, 0, 0, 1); e^\mu_{(1)} =
(0 ,1, 0, 0) ; e^\mu_{(2)} = (0 , 0, 1, 0)$.

Then, the  function (22) is the eigenfunction of the operators
${\hat p}^1$,  ${\hat p}^2$and ${\hat p}^0 - {\hat p}^3$ which
eigenvalues are $p^1$ and $p^2$ and $p^0- p^3$, respectively.
Thereat, the operators ${\hat p}^1$,  ${\hat p}^2$and ${\hat p}^0
- {\hat p}^3$ commutate with the Hamiltonian ${\cal H}$:

\begin{widetext}
\begin{eqnarray}
&& {\cal H} = \gamma^0 {\vec \gamma} \left(  {\hat {\vec p}} - g
T^a {\vec A}_a  \right) + \gamma^0 m + g A^0; \ \ \ \ \ \ A^\mu_a
= (A^0 ; {\vec A}),
\end{eqnarray}
\end{widetext}

This means that the the combinations  $P^1 = p^1; \ P^2 = p^2 ; \
P^3 = p^0 - p^3 $ of the components of the  vector $p^\mu = (p^0 ;
{\vec p})$ introduced by Eq.(17) are the quantum numbers of the
solutions  of the Dirac equation (14).

As for the spinor $v_\alpha $, we determine it by the relations:

\begin{widetext}
\begin{eqnarray}
&& v^\dag_\alpha  \ v_\beta = \delta_{\alpha \beta} ; \ \ \ \  Tr
( T_a ) = 0 ; \ \ \ \ \ Tr ( T_a \ T_b ) = {1\over 2} \delta_{a b}
\end{eqnarray}
\end{widetext}

Then, the function (22) can be  normalized by the
$\delta$-function as follows:

\begin{widetext}
\begin{eqnarray}
&& \int d^3 x  \Psi_{\sigma, \alpha}^\ast (x, p^\prime )
\Psi_{\sigma, \alpha} (x, p)   = (2\pi )^3 \delta^3 ( {\vec p} -
{\vec p}^{\ \prime} ).
\end{eqnarray}
\end{widetext}

  Direct calculations show that $\Phi_{\sigma, \alpha} (x, p)$ and $\Phi_{- \sigma,
\alpha} (x, - p)$ are orthogonal. In this way, it is obvious, that
$\Phi_{\sigma, \alpha} (x, p)$ is the so-called  positively
frequency function whereas $\Phi_{ - \sigma, \alpha} (x, - p)$ is
negatively frequency one\cite{16,17}. This fact allows us to
construct the general solution of the Dirac equation which
describes  the states both particles and anti-particles. Since the
quantum number  $P^1 = p^1; \ P^2 = p^2 ; \ P^3 = p^0 -  p^3 $ are
linearly related to the vector ${\vec p} $  we can derive the
general solution of the Dirac equation by combining the functions
$\Phi_{\sigma, \alpha} (x, p)$ and $\Phi_{- \sigma, \alpha} (x, -
p)$. As a result, due to the completeness condition (26)  the
general solution of Eq.(14)  is

\begin{widetext}
\begin{eqnarray}
&& \Psi (x) = \sum\limits_{\sigma ,  \alpha } \int {d^3 p \over
\sqrt{2p^0 \ } (2\pi)^3 } \left\{ {\hat a}_{\sigma , \alpha}
({\vec p}) \Psi_{\sigma , \alpha }  ( x, p ) +
{\hat b}^\dag_{\sigma , \alpha } ({\vec p}) \Psi_{ - \sigma , \alpha }  ( x, - p )  \right\} \nonumber \\
&& {\bar \Psi (x)}  = \sum\limits_{\sigma ,  \alpha} \int {d^3 p
\over \sqrt{2p^0 \ } (2\pi)^3 } \left\{ {\hat a}^\dag_{\sigma ,
n\alpha} ({\vec p})\ {\bar \Psi}_{\sigma , \alpha }  ( x, p ) +
{\hat b}_{\sigma , \alpha} ({\vec p}) \ {\bar \Psi}_{ - \sigma ,
\alpha } ( x,
 - p ) \right\} ,
\end{eqnarray}
\end{widetext}
where the symbols ${\hat a}^\dag_{\sigma , \alpha} ({\vec p})
;{\hat b}^\dag_{\sigma , \alpha} ({\vec p})$ and ${\hat a}_{\sigma
, \alpha} ({\vec p}); {\hat b}_{\sigma , \alpha} ({\vec p})$ are
the operators of creation and cancellation of a fermion (${\hat
a}_{\sigma , \alpha} ({\vec p}) ;{\hat a}^\dag_{\sigma , \alpha}
({\vec p})$) and anti-fermion (${\hat b}_{\sigma , \alpha} ({\vec
p}) ;{\hat b}^\dag_{\sigma , \alpha} ({\vec p})$) ,
respectively\cite{22,23}. Thereat,  ${\hat a}_{\sigma , \alpha}
({\vec p})$ ¨ $ {\hat a}^\dag_{\sigma , \alpha} ({\vec p})$;
${\hat b}_{\sigma , \alpha} ({\vec p})$ and $ {\hat
b}^\dag_{\sigma , \alpha} ({\vec p})$ satisfy the standard
commutative relations for the fermion operators.

\section{Solution of YM equation in the eikonal  approximation}

Let us fix the state of the fermion vacuum  so that the bilinear
combinations of the operators of creation and cancellation of
fermions are diagonal.  Substituting ${\bar \Psi (x)}$ and $ \Psi
(x)$ given by Eq.(22), (27) into the formula (3) we derive the
following after direct calculations:

\begin{widetext}
\begin{eqnarray}
&& J^\nu_a \ = \ {\bar \Psi} (x) \gamma^\nu T_a  \Psi(x) \ \  =
\cos^2 \theta \
 \  \sum\limits_{\sigma \alpha }\int {d^3 p
\over  p^{(0)} (2\pi)^3 } \Bigg\{  - g  A^\nu_a \
\ {\tan \theta \over  2 N \theta} - \nonumber \\
&&   g^3   { \theta \tan^2 \theta + \theta - \tan \theta \over 4
N^2 (p k )^2 \theta^3 } \int\limits_0^\varphi d \varphi^\prime
\left( A^\mu_a ( \varphi^\prime ) \  p_\mu \right)
 \left( A^\nu_b (\varphi ) \ \int\limits_0^\varphi d
\varphi^\prime \left( A_\mu^b ( \varphi^\prime ) \  p^\mu \right)
\right) + {g k^\nu \over ( p k)} \cdot \nonumber \\
&& \Bigg[ \ {\tan \theta \over  2 N \theta}  \left( A^\mu_a (
\varphi ) \  p_\mu \right) + g^2 { \theta \tan^2 \theta + \theta -
\tan \theta \over  4 N^2 (p k )^2 \theta^3 } \int\limits_0^\varphi
d \varphi^\prime \left( A^\mu_a
( \varphi^\prime ) \  p_\mu \right) \nonumber \\
&&  \left(  \left( A^\mu_b ( \varphi ) \  p_\mu \right) \
\int\limits_0^\varphi d \varphi^\prime \left( A_\mu^b (
\varphi^\prime ) \  p^\mu \right) \right) \Bigg] \Bigg\} \ \langle
{\hat a}^\dag_{\sigma , \alpha} (p) {\hat a}_{\sigma , \alpha} (p)
+ {\hat b}_{\sigma , \alpha} (p) {\hat b}^\dag_{\sigma , \alpha}
(p) \rangle  \nonumber \\
\end{eqnarray}
\end{widetext}
where the angle brackets mean  averaging over the vacuum state of
 fermions.

When the fermion system is homogeneous and isotropic the integrals
containing the square bracket are equal to zero owing to the
relativistic invariance  (see Appendix B). Then, we substitute the
current given by Eq.(28) and the field $A^\nu_a$ governed by
Eq.(13) into Eq.(20). In taking into account of Eqs.(10)¡ (20)
(that leads to cancellation of the first two terms in the
left-hand side of Eq.(12)), we derive:

\begin{widetext}
\begin{eqnarray}
&& 2 f_{ab}^{\ \ c} \sin \left(\varphi_b - \varphi_c \right) =
 f_{ab}^{\ \ c}\left\{ \ f_{c}^{\ \ sr}  \cos \left(
\varphi_b - \varphi_r \right) + \left\{ \cos \left( \varphi_b -
\varphi_r \right) \  \cos \left( \varphi_s - \varphi_a \right)
\right\} {f_{c}^{\ \ bs }  \over  N } \right\} {\cal B}_s  ;
\end{eqnarray}
\end{widetext}

\begin{widetext}
\begin{eqnarray}
&&  A^2 \cdot C    = - (N^2 -1)\sum\limits_{\sigma \alpha }\int
{d^3 p \over p^{(0)} (2\pi)^3 } \langle {\hat a}^\dag_{\sigma ,
\alpha} ({\vec p}) {\hat a}_{\sigma , \alpha} ({\vec p}) + {\hat
b}_{\sigma , \alpha} ({\vec p}) {\hat b}^\dag_{\sigma , \alpha}
({\vec p}) \rangle ,
\end{eqnarray}
\end{widetext}
where
\begin{widetext}
\begin{eqnarray}
&& C =  \  f_{ab}^{\ \ c}\ f_{c}^{\ \ sr} \left\{ \cos \left(
\varphi_b - \varphi_r \right) \  \cos \left( \varphi_s - \varphi_a
\right) \right\} < 0 .
\end{eqnarray}
\end{widetext}

The equations (29), (30) are  closed with respect to the unknown
quantities  $A$ and ${\cal B}_a$. Having been solved they
determine both the fermion and gauge  field  by means of Eqs.(13),
(22), (27) so that the wave surface $\varphi (x)$ is governed by
the  relations (10), (19), (20).

Note that in the case of  the $N=2$ (when the $SU(2)$ gauge
symmetry occurs) the convolution (31), containing cosines, always
is positive since the structure constants $f_{ab}^{\ \ c}$ are the
completely antisymmetrical tensor of the third rang
$\varepsilon_{ab}^{\ \ c}$ due to the Jacob equality\cite{22}:

\begin{widetext}
\begin{eqnarray}
&& C =  \  f_{ab}^{\ \ c}\ f_{c}^{\ \ sr} \left\{ \cos \left(
\varphi_b - \varphi_r \right) \  \cos \left( \varphi_s - \varphi_a
\right) \right\} = \sum\limits_{a,b=0}^{N^2 - 1} \sin^2 \left(
\varphi_s - \varphi_a \right) \geq 0.
\end{eqnarray}
\end{widetext}

This means that in  the framework of the developed model there is
no self-consistent solution of the Dirac and Yang-Mills equations
in the case of the $SU(2)$ gauge symmetry. When the group
dimension is more then $N=2 $ the structure constants $f_{ab}^{\ \
c}$ can not be expressed in terms of the tensor
$\varepsilon_{ab}^{\ \ c}$. As a result, it possible to fix the
differences between phase in the convolution $C$ so that $C\le 0$.

As for the coefficients ${\cal B}_s$ they  satisfy  the set of
linear algebraical equations. The matrix of this set is
symmetrical and, moreover, its diagonal elements are not all equal
to zero. This means that the equation for ${\cal B}_s$ has the
unambiguous solution.

 As a result, we have the following. The problem governed by Eqs.(1)-(5)
has the unique solution when $N\geq 3$.  The solutions are
determined by Eqs.(13), (27), (29), (30)  and correspond to the
eikonal consideration when the wave surface of the fields are
determined by the equations:

\begin{widetext}
\begin{eqnarray}
&& \left( \partial_\mu  \varphi (x) \right) \cdot \left(
\partial^\mu  \varphi (x) \right) = 0 ;\ \ \ \  \left( \partial_\mu  \partial^\mu \right) \varphi (x)  = 0
\end{eqnarray}
\end{widetext}

It follows from Eqs. (13), (27), (29), (30) that the Yang-Mills
and Dirac equations have the self-consistent solution when the
fermion current compensates the current of the gauge field which
takes place due to self-interaction of such field. In other words,
in the the framework of the developed model there is no the YM
field without fermions. In terms of QCD this means  that quarks
and gluons can not exist separately in such approach.

We should note here that the second relation in Eq.(33) implies
that the function $\varphi (x)$ which is the argument in the
expansion (13) of the field $A^\nu_a$ is the so called harmonic
function. Owing to the initial conditions it can be always  taken
such that the field $A^\nu_a$ will be localized in the confined
region of space. In this way, the relations (10),  (33) and (13)
directly lead to the axial gauge (11).

\section{Quantizing the Yang-Mills field}

We note, that  the formula (13) determining the YM field can  be
rewritten as follows:

\begin{widetext}
\begin{eqnarray}
&& A^\nu_a (x) ={1\over \sqrt{V} \sqrt{2\omega}}\sum\limits_{{\vec
q} , \alpha }   \ \left\{ c_{\alpha} ({\vec q}) e^\nu \  \exp (-i
\varphi_a ) \ \exp ( -i q x ) + c_{\alpha}^\ast ({\vec q})
{e^\ast}^\nu  \ \exp ( i \varphi_a ) \ \exp ( i q x
) \right\} + {\cal B}_a k^\nu \nonumber \\
&& c_{\alpha} ({\vec q}) =  A \sqrt{V} \sqrt{2\omega} \int {d^3
{\vec x} \over 2 (2\pi
)^3 } \exp (i \ {\vec  q}\  {\vec x} - i \varphi (x)  ) ;\nonumber \\
&& \ \ e^\nu e_\nu = {e^\ast}^\nu e^\ast_\nu = 0 ; \ \ \ e^\nu
e^\ast_\nu = 1 ; \ \ q =  (\omega ; {\vec q} ) ; \ \ q^\mu q_\mu =
0,
\end{eqnarray}
\end{widetext}
where $\alpha =1 \div 2$ takes into account of two polarizations
of the YM field, $V$ is the normalizing volume.

 Let us change the coefficients $c_{\alpha} ({\vec q})$ and
$c_{\alpha}^\ast ({\vec q})$ by the operators of cancellation
${\hat c}_{\sigma}( {\vec q}) $ and creation ${\hat
c}_{\sigma}^\dag( {\vec q}) $ of a quant of the YM field so that
${\hat c}_{\sigma} ({\vec q})$ and ${\hat c}_{\sigma}^\dag ({\vec
q})$ satisfy the Bose-Einstein commutative relation. In the
quasi-classical approximation (19) this  means that ${\hat
c}_{\sigma}( {\vec q}) $ and ${\hat  c}_{\sigma}^\dag( {\vec q}) $
commutate each other:

\begin{widetext}
\begin{eqnarray}
&& {\hat c}_{\sigma}( {\vec q}) \ {\hat c}_{\sigma}^\dag( {\vec
q}) \approx  {\hat c}_{\sigma}^\dag( {\vec q}) \  {\hat
c}_{\sigma}( {\vec q}) ,
\end{eqnarray}
\end{widetext}
where $\sigma$ is the spin variable.

Taking  into account of Eqs.(10) and (20), we  substitute Eq.(34)
into the formula (12). As a result,   we derive the relation
between the occupancy number of fermions and YM quants:

\begin{widetext}
\begin{eqnarray}
&&   \ {C\over 2 } \ \sum\limits_{{\vec q}, {\sigma} } {1\over 2
\omega ({\vec q})} \ \left\{ c_{\sigma}^\dag ({\vec q}) \
c_{\sigma} ({\vec q}) \right\} = - (N^2 -1) \   \
\sum\limits_{\sigma ; \alpha ; \ {\vec p} }  {1 \over
p^{(0)}({\vec p} )} \left\{ {\hat a}^\dag_{\sigma , \alpha} ({\vec
p}) {\hat a}_{\sigma , \alpha} ({\vec p}) + {\hat b}_{\sigma ,
\alpha} ({\vec p}) {\hat b}^\dag_{\sigma , \alpha} ({\vec p})
\right\} ,
\end{eqnarray}
\end{widetext}
where all notations are the same as they are in Eq.(30), (31),
(34).

Eqs.(27), (35) allow us to obtain  the energy $E$ of interacting
fermions. Calculating the energy-momentum tensor $T^{\nu \mu}$ we
derive

\begin{widetext}
\begin{eqnarray}
&&   E = \sum\limits_{{\vec q}, {\sigma} } \omega ({\vec q}) \
\left\{ c_{\sigma}^\dag ({\vec q}) \ c_{\sigma} ({\vec q})
\right\} + \ \sum\limits_{\sigma ; \alpha ; \ {\vec p} } \ p^{(0)}
({\vec p}) \left\{ {\hat a}^\dag_{\sigma , \alpha} ({\vec p})
{\hat a}_{\sigma , \alpha} ({\vec p}) + {\hat b}_{\sigma , \alpha}
({\vec p}) {\hat b}^\dag_{\sigma , \alpha} ({\vec p}) \right\} .
\end{eqnarray}
\end{widetext}

It follows from the last expression that  the energy of the system
of particles which consists of  fermions and gauge quants splits
on two terms. However such additivity is fictitious since the
occupancy numbers of the fermion field depend on the value of the
YM field.    Formally, this manifests itself via Eq.(36) at the
quasi-classical level.

\section{Dirac and Yang-Mills Fields as strong interacting matter}

Let us  consider the solution of Eq.(29)-(30)  in detail,  when
$N\geq 3$. We assume that the phases $\varphi_a$ are chosen so
that the convolution $C$ given by Eq.(31) is negative.

\subsection{Fermion fields}

In principle,  Eqs. (13), (27), (30) allow  us to calculate  both
the amplitude and phases of the YM field provided that  the
occupancy number of fermions  is known. However, the relation (30)
is some functional equation since the right-hand side of it
depends on the required value $A$ via the 4-momentum $p^\mu = (
E({\vec p}) ; {\vec p}) $ of a fermion  in an external field (see
Eq.(27)) which enters into the correlators of fermi-operators.
Such complicate problem is simplified and can be solved in a very
important case when the system of fermions is some equilibrium
matter whose temperature is $T$. Then, the correlators in Eq.(30)
are the equilibrium  occupancy numbers $n(E)$ which are equal to
\cite{26}:

\begin{widetext}
\begin{eqnarray}
&& \langle {\hat a}^\dag_{\sigma , \alpha} (p) {\hat a}_{\sigma ,
\alpha} (p) \rangle = \langle {\hat b}_{\sigma , \alpha} (p) {\hat
b}^\dag_{\sigma , \alpha} (p) \rangle = \cong {1\over 1 +  \exp
\left( { p^\mu U_\mu - \ \mu \over T }  \right) } \cong {1\over 1
+ \exp \left( { Q^\mu U_\mu - \ \mu \over T }  \right) } \equiv
n(Q^0 ),
\end{eqnarray}
\end{widetext}
where  $\mu$ is the chemical potential supposed to be the same for
all type of fermions;  $Q^\mu$ is the mean value of a kinetic
momentum of a fermion, $Q^\mu = p^\mu - g T^a A_a^\mu$, in an
external field; $U_\mu = (1, 0,0,0)$ is the so-called
hydrodynamics velocity\cite{26}.

The functions (22) allow us to drive the mean value  of the
kinetic momentum  of a fermion $Q^\mu$. After a direct
calculations we get

\begin{widetext}
\begin{eqnarray}
&&  (Q^0 )^2 = {\vec p}^{\ 2} + m^2_\ast ; \ \ \ \ m_\ast^2 = m^2
+ {g^2 (N^2 - 1 ) A^2 \over 2 N}
\end{eqnarray}
\end{widetext}

Since  $Q^0$ can be interpretable  as the mean value of the energy
$E$ of a fermion in an external field, the last equation means
that the interaction of a fermion with an  external field leads to
the re-normalization of a  fermion mass, in the mean.

Substituting Eqs.(27), (38)  into the formula (30),  we obtain:

\begin{widetext}
\begin{eqnarray}
&&  A^2 \cdot\vert C \vert    = 2  N (N^2 - 1) \left( {T\over
\pi}\right)^2 \int\limits_{z}^{\infty} {\sqrt{x^2 - z^2 }\ d x
\over 1 + \exp \left( x - {\mu\over T}\right)},\ \ \ \ \ \ z =
{m_\ast \over T} = {1\over T} \sqrt{  m^2 + {g^2 (N^2 - 1 ) A^2
\over 2 N}},
\end{eqnarray}
\end{widetext}

To derive the solution of Eq.(40), first,   we assume that the
occupancy numbers are not too large so that the Boltzmann
distribution is applicable to fermions.  In this case the chemical
potential can be expressed via the density of particles $n_0$
\cite{26}. Then, we obtain from Eq.(40) at $T\gg m$ :

\begin{widetext}
\begin{eqnarray}
&&  z^3 \ K_2 (z) = {n_0 \over \vert C\vert  T^3 }  \ K_1(z); \ \
\ \ \ \ T\gg m ,
\end{eqnarray}
\end{widetext}
where $K_\nu (z)$ are  the modified Bessel functions (the McDonald
functions)\cite{27} .

The ratio $K_1 (z) / K_2 (z)$  monotonically increases,  so
that\cite{27}:

\begin{widetext}
\begin{eqnarray}
&& K_1 (z) / K_2 (z) = z/2 ; \ \ \ \ \  z\to 0 \nonumber \\
&& K_1 (z) / K_2 (z) = 1 ; \ \ \ \ \  z\to \infty .
\end{eqnarray}
\end{widetext}

Then, Eq.(41) has the unambiguous solution which is

\begin{widetext}
\begin{eqnarray}
&& z = \left( {n_0 \over 2 \vert C\vert  T^3 } \right)^{1/2} ; \ \
\
\ \ z\ll 1 \nonumber \\
&& z = \left( {n_0 \over  \vert C\vert  T^3 } \right)^{1/3} ; \ \
\ \ \ z\gg 1
\end{eqnarray}
\end{widetext}

Since the Bolzmann approximation is correct when $(n_0  /T^3 ) \ll
1$, the first formula in Eq.(43) can only  be used for calculation
of the amplitude $A$ of the YM field. Then,  we derive from
Eq.(40):

\begin{widetext}
\begin{eqnarray}
&&   A = {2 N \over  g (N^2 - 1 )} \left( {n_0 \over 2 \vert
C\vert T^3 } \right)^{1/2} \ T ; \ \ \ \ \ {m\over T} \ll  \left(
{n_0 \over  \vert C\vert T^3 } \right) \ll 1 ; \nonumber \\
\end{eqnarray}
\end{widetext}

It follows from Eq.(44) that at $(n_0  T^{-3}) \lesssim 1$  the
field amplitude $A$ is such that  the effective mass of a fermion
is small, $m_\ast \ll T $, even in the presence of the external
field, i.e. the fermions in the external field remain  an
ultrarelativistic particles as before.

On the other hand, the wave packet of an ultrarelativistic $ E\gg
m $ particle does not spread out as compared with the case of
non-relativistic particles owing to the dispersion law which is
$E(p)\ \sim p \ $ at $ E\gg m $ . This means that fermion states
in the external field are single-particle ones in this case. Then,
the system  of fermions can be considered as some matter
consisting of individual fermi-particles so that the interaction
of them with the field results in the renormalization of their
masses, in the mean. In this way, the individuality of a particle
keeps as soon as the density of the matter is not too large $(n_0
T^{-3}) \lesssim 1$\cite{28}.

With  increasing the number of fermions (or with  decreasing the
matter temperature) the Boltzmann approach becomes unsuitable. In
the case of $n_0 \gtrsim T^3$, the chemical potential $\mu$ is of
the order of $\mu \sim n^{2/3}_0 \gg T$. Then, transforming the
integral in Eq.(40) according to Ref.\cite{28},  we obtain:

\begin{widetext}
\begin{eqnarray}
&&   A = \left( {2 N \over  g (N^2 - 1 )} \right)^{1\over 2} \ \mu
\simeq \left( {2 N \over  g (N^2 - 1 )} \right)^{1\over 2} \
n^{2\over 3} _0 \gg  \ T ; \ \ \ \ \ {\mu \over T} \gtrsim  z  \gg
1 .
\end{eqnarray}
\end{widetext}

The last formula shows  that in decreasing the temperature of
matter (or in increasing its density) the mean effective mass of a
fermion  $m_\ast$ in the external field  is enlarged, so that  the
fermions become non-relativistic  particles. That leads to
delocalizing the fermion states in the space due to spreading out
of a wave function. Thus, in this case the matter constitutes some
fermi-liquid consisting of fermions with the renormalized mass.

\subsection{Yang-Mills fields}

The obtained  formulae (29), (30) allow us to get the tensor
$F^{\nu \mu}_a (x)$ of the gauge YM field. After  direct
calculations we get from Eqs.(2), (13):

\begin{widetext}
\begin{eqnarray}
&& F^{\nu \mu}_a =( \partial_\nu \varphi ) {\partial A^\mu_a \over
\partial \varphi} -
( \partial_\mu \varphi ) {\partial A^\nu_a \over \partial
\varphi}.
\end{eqnarray}
\end{widetext}

The derived tensor $F^{\nu \mu}_a$ enables us to obtain the
strength of both a "electric" and "magnetic"  field as well as the
energy-momentum tensor $T^{\nu \mu}$. The diagonal components of
$T^{\nu \mu}$ give the energy density $w$ and Pointing vector
${\vec S}$:

\begin{widetext}
\begin{eqnarray}
&& T^{\nu \nu} = ( \partial^\nu \varphi ) \cdot  ( \partial^\nu
\varphi )\ A^2  \equiv ( w ; {\vec S} ).
\end{eqnarray}
\end{widetext}

The last formulae show that there is no  energy flux of the gauge
field through any  surface confining the range where the YM field
is. This means that the role of the YM field in the considered
approximation is binding fermion what leads to the renormalization
of their masses.

\section{Developed model in context of QCD}

\subsection{ Effective mass of quarks}

First, we discuss  applicability of the developed model to
description of the strong interacting matter generated in
collisions of heavy ions of high energies. The quasi-classicality
of the model means that the occupancy number of particle are
large.

In the RHIC and SPS experiments the characteristic temperature $T$
of an equilibrium quark-gluon plasma is  $T\sim 200 \div 400 MeV$.
The estimations of the initial density of energy of the plasma
give that the energy density $w \sim 10  \ Gev \cdot F^{-3}$ while
the volume of the fireball is not less than $ V_0 \sim 10^2 \
F^3$. Then  the number of particles $N$ inside the fireball  is of
the order of

\begin{eqnarray}
&& N\sim {w \ V_0 \over T } \gtrsim 2.5 \cdot 10^3 ,
\end{eqnarray}
that is in agreement with the quasi-classical approximation.

 The gas parameter $n_0^{1/3}
T^{-1} $ is of the order of $( n_0^{1/3} T^{-1}) \sim  1.46 \div
3.7$ at such  density  of the matter. On the other hand,  the mean
effective  mass of a quark  is of the order of

\begin{eqnarray}
&& m_\ast \sim   \left( {2 N \over  g (N^2 - 1 )} \right)^{1\over
2} \ \left( {n_0 \over 2 \vert C\vert T^3 } \right)^{1/2} \ T \  ;
\ \ \ \ \ {m\over T} \ll  \left( {n_0 \over  \vert C\vert T^3 }
\right)
 \ll 1 ; \nonumber \\
 && m_\ast \sim   \  \sqrt[3]{n_0} \  ; \ \ \ \ \ \left( {n_0 \over  \vert C\vert T^3 } \right)
 \gg 1 .
\end{eqnarray}

It follows from the last formulae that in the intermediate range
of the density of matter  $n_0 \sim (g T)^3 $ the effective mass
is proportional to the temperature of the matter that corresponds
to the result of the calculation of the thermal mass of a quark in
the hard loop approximation\cite{29,30}:

\begin{eqnarray}
m_\ast \sim  g \ T .
\end{eqnarray}

\subsection{Hadronization}

Arising  the collective fermi-liquid states of fermions, which are
governed  by Eqs.(40), (45), is typical in the situation when
there are no any   channel for the particles to escape the fermion
system. In the case of a quark-gluon plasma quarks can go out of
the system  due the process of the hadronization.

We  estimate the mass of the hadron generated  in the result of
the hadronization. We assume that the hadronization is the
equilibrium phase transition of the first kind. Then,  the
chemical potentials of the quarks $\mu_q$, gluons $\mu_g$ and
hadrons $\mu_h$ are:

\begin{eqnarray}
\mu_q = \mu_g = \mu_h =0
\end{eqnarray}

Substituting $\mu_q =  0$ into Eq.(40) we derive:

\begin{eqnarray}
 A \sim 1.5 \ T_c ; \ \ \ \  \ m_{\ast} \sim 1.78 \  T_c  ,
\end{eqnarray}
where $T_c$ is the temperature of the considered phase transition.
In obtaining Eqs.(52),   we set  $g=1; \ \vert C\vert \simeq 1$.

Let us the pions  $\pi^0$  are only created in the result of the
phase transition. Since $\pi^0$'s consist of $12$ quarks
(including anti-quarks) we equalize the number of hadrons and
quarks (divided by $12$) at the temperature $T_c$. As a result, we
obtain:

\begin{eqnarray}
{3\over 2} \int\limits_0^{\infty} {x^2 dx \over \exp\left(
\sqrt{x^2 + (M_h / \ T_c )^2 } \right) - 1 } =
\int\limits_0^{\infty} {x^2 dx \over \exp\left( \sqrt{x^2 + (1.78
)^2 }  \right) + 1 } ,
\end{eqnarray}
where $M_h$ is the hadron mass. In deriving the last equation we
take into account  that the effective  mass of a quark is
$m_{\ast} \sim 1.78 \ T_c$ according to Eqs.(52).

Solution of Eq.(53) gives that $M_h \simeq 0.68 \ T_c $. If the
temperature of the phase transition is $T_c \simeq 200 MeV$, the
hadron mass is $M_h \simeq 136 Mev$, that corresponds to the mass
of a free pion.

\section{Conclusion}

The quasi-classical  model in the gauge $SU(N)$ field theory is
considered when the YM field is assumed to be in the form of the
eikonal wave. The  self-consistent solutions of the
non-homogeneous YM equation and  the Dirac equation in the
external YM field is derived. It is shown that  the considered
problem is solvable when the dimension of the gauge group $N\geq
3$. Thereat, the currents generated by fermions and gauge field
exactly compensate each other.

In terms of the multi particle problem,  the obtained solutions
correspond to the both individual and collective states of
fermions in matter that depends strongly on the parameters of the
problem such as the density of fermions and the  temperature of
matter. As for the YM field,  its amplitude appears to depend
strongly on the number of fermions so that the field  does not
exist without fermions. The derived gauge field has the form of a
circulatory polarized  wave (see Eqs.(13)) which energy is
concentrated in  the localized region of  space. Thereat,
interaction of the  YM field with  fermions leads, in the mean, to
the re-normalization of a fermion mass so that it  enlarges with
increasing  the YM field amplitude .

The quantum theory of the considered model is developed in the
quasi-classical approximation. The energy of the quantized fields
is obtained . It shown that the energy strongly depends on the
derived relation between the occupancy numbers of fermions and
quants of the YM field.

The relation of the developed model to the generally accepted
results in QCD is considered.  In the case of the hot homogeneous
equilibrium quark-gluon plasma  the re-normalization of a fermion
mass leads to arising the thermal mass of a quark (see Eqs.(40),
(49)) which strongly depends on the matter temperature. We show
that in the intermediate range of the density and temperature of
the plasma, $n_0 \sim g^3 T^3 $, the dependence of the quark mass
on the matter temperature and coupling constant (see Eq.(50))
corresponds to the results of it calculations which have been made
in the hard thermal loop approximation\cite{29,30} early. The
hadronization as the phase transition of the first kind is
considered. In the case of the hadronization into the  lightest
hadrons the calculated  mass of such hadrons appears to be of the
order of the mass of  a free pion.

\appendix
\section{}

We expand the second exponent in Eq.(20) in the series:

\begin{widetext}
\begin{eqnarray}
&&  \exp \left\{ - i g \ T_a { \int\limits_0^\varphi d
\varphi^\prime \left( A_\mu^a p^\mu \right)\ + {i\over 2}
\left(\gamma^\mu k_\mu
 \right) \left( \gamma^\mu
 A_\mu^a  \right) \over  ( p k ) } \right\}  \equiv
 \exp \left\{ - { i g  \over  ( p k ) } \ T_a  B^a ( \varphi )
\right\} = \nonumber \\
&& 1 + \left( - { i g  \over  ( p k ) } \right) \left( T_a  B^a
\right) + {1\over 2 !} \left( - { i g \over ( p k ) } \right)^2
\left( T_a  B^a \right)^2 + {1\over 3 !} \left( {- i g \over ( p k
) } \right)^3 \left( T_a  B^a \right)^3 + {1\over 4 !} \left( - {
i g \over ( p k ) } \right)^4 \left( T_a  B^a \right)^4 + \dots
\nonumber \\
&& B^a (\varphi ) \equiv  \int\limits_0^\varphi d \varphi^\prime
\left( A_\mu^a ( \varphi^\prime ) \  p^\mu \right) \  + {i\over 2}
\left( k_\mu  \gamma^\mu \right) \left( A_\mu^a ( \varphi ) \
\gamma^\mu \right); \ \ \ B^a (\varphi ) B_a (\varphi ) =
 \left(
\int\limits_0^\varphi d \varphi^\prime \left( A_\mu^a (
\varphi^\prime ) \  p^\mu \right) \ \int\limits_0^\varphi d
\varphi^\prime \left( A_a^\mu ( \varphi^{\prime \prime }) \ p_\mu
\right) \right) \nonumber \\
&& + i  \left( k_\mu  \gamma^\mu \right) \left( A_\mu^a ( \varphi
) \ \gamma^\mu \right) \left( \int\limits_0^\varphi d
\varphi^\prime \left( A^\mu_a ( \varphi^\prime ) \  p_\mu \right)
\right) \equiv \left( c^2 + b \right)
\end{eqnarray}
\end{widetext}

Taking into account  of Eqs.(6)-(8), (10), (11)  the even and odd
terms in Eq. (A.1) can be rewritten as follows:

\begin{widetext}
\begin{eqnarray}
&&  \left( \exp \left\{ - i g \ T_a { \int\limits_0^\varphi d
\varphi^\prime \left( A_\mu^a p^\mu \right)\ + {i\over 2}
\left(\gamma^\mu k_\mu
 \right) \left( \gamma^\mu
 A_\mu^a  \right) \over  ( p k ) } \right\} \right)_{even}
  =  1 -  {1\over 2 !} \left( {  g  \over ( p k )  } \sqrt{{N^2 -
2\over 4N}} \right)^2 \left( c^2 + b \right) + \nonumber \\
&& {1\over 4 !} \left( { g \over ( p k )  } \sqrt{{N^2 - 2\over
4N}} \right)^4 \left( c^2 + 2 c^2 b \right) - {1\over 6 !} \left(
{  g  \over ( p k )  } \sqrt{{N^2 - 2\over 4N}} \right)^6 \left(
c^6 + 3 b c^4 \right) + \dots = \nonumber \\
&& \cos \theta - {b\over 2} \left( { g  \over ( p k )  }
\sqrt{{N^2 - 2\over 4N}} \right)^2  \ {\sin \theta \over \theta }
\end{eqnarray}
\end{widetext}

\begin{widetext}
\begin{eqnarray}
&& \left( \exp \left\{ - i g \ T_a { \int\limits_0^\varphi d
\varphi^\prime \left( A_\mu^a p^\mu \right)\ + {i\over 2}
\left(\gamma^\mu k_\mu
 \right) \left( \gamma^\mu
 A_\mu^a  \right) \over  ( p k ) } \right\} \right)_{odd}
  = - {i q T_a B^a \over ( p k ) } \Bigg(  1 -  {1\over 3 !} \left( {  g  \over ( p k )  } \sqrt{{N^2 -
2\over 4N}} \right)^2 \left( c^2 + b  \right) + \nonumber \\
&& {1\over 5 !} \left( { g \over ( p k )  } \sqrt{{N^2 - 2\over
4N}} \right)^4 \left( c^2 + 2 c^2 b \right) - {1\over 7 !} \left(
{ g  \over ( p k )  } \sqrt{{N^2 - 2\over 4N}} \right)^6 \left(
c^6 + 3 b c^4 \right) + \dots \Bigg) = \nonumber \\
&& - {i q T_a B^a \over ( p k ) } \left( {\sin \theta \over \theta
} + {b\over 2} \left( { g  \over ( p k )  } \sqrt{{N^2 - 2\over
4N}} \right)^2  { \theta \cos \theta -  \sin \theta \over \theta^3
 }\right) ,
\end{eqnarray}
\end{widetext}
where the parameters $c$ and $b$ are determined in Eq.(A1).

Summing  the exponents in Eq.(A.2), (A.3) we directly go to
Eq.(21).

\section{}

Due to the relativistic invariance and Eqs.(10), (11) the integral
containing the first term in the square bracket in Eq.(28) is
equal to zero:

\begin{widetext}
\begin{eqnarray}
&&   \sum\limits_{\sigma \alpha }\int {d^3 p \over 2 p^{(0)}
(2\pi)^3 } \Bigg\{  {g k^\nu \over ( p k)} \cdot  \Bigg[ {N^2
-2\over 4N} \ {\tan \theta \over \theta}  \left( A^\mu_a ( \varphi
) \  p_\mu \right)  \Bigg] \Bigg\} \ \langle {\hat a}^\dag_{\sigma
, \alpha} (p) {\hat a}_{\sigma , \alpha} (p) + {\hat b}_{\sigma ,
\alpha} (p) {\hat b}^\dag_{\sigma , \alpha}
(p) \rangle =  \nonumber \\
&& g k^\nu \ {N^2 -2\over 4N}  A^\mu_a ( \varphi )
\sum\limits_{\sigma \alpha }\int {d^3 p \over 2 p^{(0)} (2\pi)^3 }
\Bigg\{  {1\over ( p k)} \cdot \Bigg[  \ {\tan \theta \over
\theta}   p_\mu
 \Bigg] \Bigg\} \ \langle {\hat a}^\dag_{\sigma , \alpha} (p)
{\hat a}_{\sigma , \alpha} (p) + {\hat b}_{\sigma , \alpha} (p)
{\hat b}^\dag_{\sigma , \alpha} (p) \rangle =
\nonumber \\
&& k^\nu f_1 (\varphi )   \left(  A^\mu_a ( \varphi ) k_\mu
\right) = 0 ,
\end{eqnarray}
\end{widetext}
where $f_1 (\varphi) $ is some scalar function.

Because of the relativistic invariance, the fact that $J^\nu_a$ is
a $SU(N)$ vector, and  Eqs.(10), (11)   the integral containing
the second term in the square bracket in Eq.(28) is also equal to
zero:

\begin{widetext}
\begin{eqnarray}
&&  \sum\limits_{\sigma \alpha }\int {d^3 p \over 2 p^{(0)}
(2\pi)^3 } \Bigg\{  {g k^\nu \over ( p k)} \cdot  \Bigg[  g^2
\left( {N^2 -2\over 4N}\right)^2  { \theta \tan^2 \theta + \theta
- \tan \theta \over (p k)^2 \theta^3 } \int\limits_0^\varphi d
\varphi^\prime \left( A^\mu_a
( \varphi^\prime ) \  p_\mu \right) \nonumber \\
&&  \left(  \left( A^\mu_b ( \varphi ) \  p_\mu \right) \
\int\limits_0^\varphi d \varphi^\prime \left( A_\mu^b (
\varphi^\prime ) \  p^\mu \right) \right) \Bigg] \Bigg\} \ \langle
{\hat a}^\dag_{\sigma , \alpha} (p) {\hat a}_{\sigma , \alpha} (p)
+ {\hat b}_{\sigma , \alpha} (p) {\hat b}^\dag_{\sigma , \alpha}
(p) \rangle =  \nonumber \\
&& g^3 k^\nu \left( {N^2 -2\over 4N}\right)^2 A_b^\rho ( \varphi )
\int\limits_0^\varphi d \varphi^\prime  A^\mu_a ( \varphi^\prime )
\ \int\limits_0^\varphi d \varphi^{\prime \prime}  A_\lambda^b (
\varphi^{\prime \prime } )\ \sum\limits_{\sigma \alpha }\int {d^3
p \over 2 p^{(0)} (2\pi)^3 } \Bigg\{  {\ p_\rho  p^\lambda  \
p_\mu \over ( p k)} \cdot  { \theta \tan^2 \theta + \theta - \tan
\theta \over (p k)^2 \theta^3 }    \
   \Bigg\} \  \nonumber \\
&& \langle {\hat a}^\dag_{\sigma , \alpha} (p) {\hat a}_{\sigma ,
\alpha} (p) + {\hat b}_{\sigma , \alpha} (p) {\hat b}^\dag_{\sigma
, \alpha} (p) \rangle =  \nonumber \\
&& k^\nu A_b^\rho ( \varphi ) \int\limits_0^\varphi d
\varphi^\prime  A^\mu_a ( \varphi^\prime ) \ \int\limits_0^\varphi
d \varphi^{\prime \prime}  A_\lambda^b ( \varphi^{\prime \prime }
)\ \left( k^\lambda k_\rho k_\mu f_2(\varphi ) + {\cal G}_{\rho
\mu} k^\lambda f_3(\varphi ) + {\cal G}^\lambda_{\mu} k_\rho
f_4(\varphi )+ {\cal G}^\lambda_{\rho } k_\mu f_5(\varphi )
\right) = 0
\end{eqnarray}
\end{widetext}
where $f_i (\varphi) $ are some scalar functions.

\end{document}